\documentclass[12pt]{article}
\usepackage{amsmath}
\usepackage[dvips]{graphicx}

\begin{document}
	
    \def\prg#1{\medskip\noindent{\bf #1}}
	\def\inn{\,\rfloor\,}                  \def\bt{{\bar\tau}}
	\def\br{{\bar\rho}}
	\def\prg#1{\medskip\noindent{\bf #1}}  \def\ra{\rightarrow}
	\def\lra{\leftrightarrow}              \def\Ra{\Rightarrow}
	\def\nin{\noindent}                    \def\pd{\partial}
	\def\dis{\displaystyle}                \def\inn{\,\rfloor\,}
	\def\grl{{GR$_\Lambda$}}               \def\Lra{{\Leftrightarrow}}
	\def\cs{{\scriptstyle\rm CS}}          \def\ads3{{\rm AdS$_3$}}
	\def\Leff{\hbox{$\mit\L_{\hspace{.6pt}\rm eff}\,$}}
	\def\bull{\raise.25ex\hbox{\vrule height.8ex width.8ex}}
	\def\ric{{(Ric)}}                      \def\tric{{(\widetilde{Ric})}}
	\def\tmgl{\hbox{TMG$_\Lambda$}}
	\def\Lie{{\cal L}\hspace{-.7em}\raise.25ex\hbox{--}\hspace{.2em}}
	\def\sS{\hspace{2pt}S\hspace{-0.83em}\diagup}   \def\hd{{^\star}}
	\def\dis{\displaystyle}     \def\ul#1{\underline{#1}}
	\def\pgt{{\scriptstyle\rm PGT}}
	\def\ul{{\underline}}
	
	\def\hook{\hbox{\vrule height0pt width4pt depth0.3pt
			\vrule height7pt width0.3pt depth0.3pt
			\vrule height0pt width2pt depth0pt}\hspace{0.8pt}}
	\def\semidirect{\;{\rlap{$\supset$}\times}\;}
	\def\first{\rm (1ST)}       \def\second{\hspace{-1cm}\rm (2ND)}
	\def\bm#1{\hbox{{\boldmath $#1$}}}      \def\un#1{\underline{#1}}
	\def\nb#1{\marginpar{{\large\bf #1}}}
	\def\diag{{\rm diag}\,}
	
	\def\G{\Gamma}        \def\S{\Sigma}        \def\L{{\mit\Lambda}}
	\def\D{\Delta}        \def\Th{\Theta}
	\def\a{\alpha}        \def\b{\beta}         \def\g{\gamma}
	\def\d{\delta}        \def\m{\mu}           \def\n{\nu}
	\def\th{\theta}       \def\k{\kappa}        \def\l{\lambda}
	\def\vphi{\varphi}    \def\ve{\varepsilon}  \def\p{\pi}
	\def\r{\rho}          \def\Om{\Omega}       \def\om{\omega}
	\def\s{\sigma}        \def\t{\tau}          \def\eps{\epsilon}
	\def\nab{\nabla}      \def\btz{{\rm BTZ}}   \def\heps{\hat\eps}
	\def\bu{{\bar u}}     \def\bv{{\bar v}}     \def\bs{{\bar s}}
	\def\te{{\tilde e}}   \def\tk{{\tilde k}}
	\def\bare{{\bar e}}   \def\bark{{\bar k}}
	\def\barom{{\bar\omega}}  \def\barg{{\bar g}}
	\def\tphi{{\tilde\vphi}}  \def\tt{{\tilde t}}
	\def\te{{\tilde e}}
	
	\def\tG{{\tilde G}}   \def\cF{{\cal F}}    \def\bH{{\bar H}}
	\def\cL{{\cal L}}     \def\cM{{\cal M }}   \def\cE{{\cal E}}
	\def\cH{{\cal H}}     \def\hcH{\hat{\cH}}
	\def\cK{{\cal K}}     \def\hcK{\hat{\cK}}  \def\cT{{\cal T}}
	\def\cO{{\cal O}}     \def\hcO{\hat{\cal O}} \def\cV{{\cal V}}
	\def\tom{{\tilde\omega}}  \def\cE{{\cal E}}     \def\tR{{\tilde R}}
	\def\cR{{\cal R}}    \def\hR{{\hat R}{}}   \def\hL{{\hat\L}}
	\def\tb{{\tilde b}}  \def\tA{{\tilde A}}   \def\tv{{\tilde v}}
	\def\tT{{\tilde T}}  \def\tR{{\tilde R}}   \def\tcL{{\tilde\cL}}
	\def\hy{{\hat y}}    \def\tcO{{\tilde\cO}} \def\hom{{\hat\om}}
	\def\he{{\hat e}}
	\def\nn{\nonumber}                    \def\vsm{\vspace{-10pt}}
	\def\be{\begin{equation}}             \def\ee{\end{equation}}
	\def\ba#1{\begin{array}{#1}}          \def\ea{\end{array}}
	\def\bea{\begin{eqnarray} }           \def\eea{\end{eqnarray} }
	\def\beann{\begin{eqnarray*} }        \def\eeann{\end{eqnarray*} }
	\def\beal{\begin{eqalign}}            \def\eeal{\end{eqalign}}
	\def\lab#1{\label{eq:#1}}             \def\eq#1{(\ref{eq:#1})}
	\def\bsubeq{\begin{subequations}}     \def\esubeq{\end{subequations}}
	\def\bitem{\begin{itemize}}           \def\eitem{\end{itemize}}
	\renewcommand{\theequation}{\thesection.\arabic{equation}}
	
\title{A black hole with torsion in 5D Lovelock gravity}
\author{ B. Cvetkovi\'c and D. Simi\'c\footnote{
		Email addresses: {cbranislav@ipb.ac.rs, dsimic@ipb.ac.rs}}\\
	Institute of Physics, University of Belgrade \\
	Pregrevica 118, 11080 Belgrade, Serbia}

\maketitle

\begin{abstract}
We analyzed static spherically symmetric solutions of the five dimensional (5D)  Lovelock gravity in the first order
formulation.  In the  Riemannian sector, when torsion vanishes,   Boulware-Deser black hole
represents a unique static spherically symmetric black hole solution  for the {\it generic} choice of the Lagrangian parameters.  We showed
that the {\it special}  choice of  the Lagrangian parameters, different from the Lovelock Chern-Simons gravity, leads to the existence of the static  black hole solution with {\it torsion},
which metric is asymptotically AdS. We  calculate conserved charges and thermodynamical quantities of the black hole solution.\\\\

\end{abstract}

\section{Introduction}

Lovelock gravity \cite{x1} represents an intriguing generalization of general relativity, since it is a unique ghost free higher derivative extension of Einstein's theory that possesses second order equations of motion.
As a higher curvature theory, Lovelock gravity has a considerable number of black hole solutions, see \cite{x4,x5,x6,x7,x8,x9,x10,x11,y1} and references therein. Many of them possess exotic properties, such as zero mass, peculiar topology of the event horizon etc.

This leads us to an old problem of black hole uniqueness, i.e. solutions of general relativity are highly constrained, but the situation changes drastically in the case   of higher dimensions. There are new black hole solutions with non-spherical topology of event horizon, namely black string, black ring and black brane \cite{x12}. Often these exotic black objects suffer from various instabilities, for example black strings and branes have Gregory-Laflamme instability \cite{x13}, and will decay into black hole with spherical horizon. Thus,  gravity in higher dimensions represents  an interesting area of research, full of surprising discoveries, whose importance stems from its numerous  applications.

Lovelock gravity can be also studied within the framework of Poincar\'e gauge theory (PGT),  formulated by Sciamma \cite{z1} and Kibble \cite{z2} more than half a century ago.  PGT is a first  modern, gauge-field-theoretic approach to gravity  obtained by gauging the Poincar\'e
group of spacetime symmetries, the semidirect product of translations and Lorentz transformations. It represents a natural extension of the  gauge principle, originally formulated by Weyl within electrodynamics  and further developed in the works of Yang, Mills and Utiyama, to the spacetime symmetries. The adopted gauge procedure leads directly to a new, Riemann-Cartan geometry of spacetime, since torsion and curvature are recovered as  the Poincar\'e gauge field strengths. The Lagrangian in PGT contains a gravitational part, which is a function of the field strengths, the curvature and the torsion, and a suitable matter field Lagrangian.

In the context of Lovelock gravity this more general setting contains torsionless theory as a limit and represents a starting point for canonical analysis, coupling with matter fields, supersymmetric extensions of the theory and holographic applications.
Interestingly, unlike in the case of Einstein-Cartan theory (first order formulation
of general relativity) where all  solutions of the equations of motion in vacuum are torsion free, the structure of the
vacuum solutions of the Lovelock gravity is more complicated, because there exist solutions with non-vanishing torsion.
  However, it turns out that exact solutions with  torsion are extremely difficult to be found, since consistency conditions usually lead to an over constrained system of equations.  Solutions with non-trivial totally antisymmetric torsion have  been studied in \cite{x10,y2,y3,y4,y5,y6}.  In this paper we continue our analysis of the exact solutions of  5D Lovelock gravity solutions with torsion, started in \cite{x10}, and find new static spherically symmetric black hole solution with torsion with zero mass and entropy. The torsion of the solution possesses both tensorial and antisymmetric part. It, unlike Riemannian  Boulware-Deser black hole \cite{x14}, exists for the specific choice of action parameters. This fine
 tuning of action parameters has been firstly noticed by F. Canfora, A. Giacomini ans S. Willinson in their paper \cite{y2} and it represents
 the sector different from the highly degenerate Lovelock Chern-Simons gravity.

The paper is organized in the following way. In the second section we review basics of   Poincar\'e gauge theory and Lovelock gravity  in the first order formulation. In section 3  we find the black hole solution of 5D Lovelock gravity with torsion and analyze its properties. In particular, we find that quadratic torsional invariant is singular at $r\ra 0$.  In section 4 we explore thermodynamics of the previously obtained solution. Appendices contain additional technical details.

We use the following conventions: Lorentz signature is mostly minus, local Lorentz indices  are denoted by the middle letters of the Latin alphabet, while space-time indexes are denoted by the letters of the Greek alphabet. Throughout the paper we mostly use differential forms instead of coordinate notation, and the  wedge product is omitted for simplicity.

\section{Lovelock gravity}
\setcounter{equation}{0}

Since the work of  Sciamma and Kibble it is known that gravity in the first order formulation has the structure of Poincar\'e gauge theory (PGT),
 see \cite{x15,x16} for comprehensive account. For the readers' convenience we briefly review basics of the PGT.

\prg{PGT in brief.} The basic dynamical variables in PGT, playing the role of gauge potentials, are the
vielbein $e^i$ 1-form and the spin connection $\om^{ij}=-\om^{ji}$ 1-form. In local coordinates $x^\m$, we can expand the vielbein and the
connection 1-forms as $e^i=e^i{_\m}dx^\m$, $\om^i=\om^i{}_\m dx^\m$.  Gauge symmetries of the theory
are local translations (diffeomorphisms) and local Lorentz rotations, parametrized by
$\xi^\m$ and $\ve^{ij}$ respectively.

From the gauge potentials we can construct field strengths, namely torsion $T^i$ and curvature $R^{ij}$ (2-forms), which are given as
\bea
&&T^i=\nabla e^i\equiv de^i+\ve^i{}_{jk}\om^j\wedge e^k
=\frac{1}{2}T^i{}_{\m\n}dx^\m\wedge dx^\n\, ,         \nn\\
&&R^{ij}=d\om^{ij}+\om^{ik}\wedge\om_k{^j}
=\frac{1}{2}R^{ij}{}_{\m\n}dx^\m\wedge dx^\n\, , \nn
\eea
where $\nabla=dx^\m\nabla_\m$ is the exterior covariant derivative.

Metric tensor can be constructed from vielbein and flat metric $\eta_{ij}$
\bea
&&g=\eta_{ij}e^i\otimes e^j=g_{\m\n}dx^\m\otimes dx^\n \, ,\nn\\
&& g_{\m\n}=\eta_{ij}e^i{_\m}e^j{_\n}\, ,
\qquad \eta_{ij}=(+,-,-)\, .                   \nn
\eea
The antisymmetry of $\om^{ij}$ in PGT is
equivalent to the so-called {\it metricity condition\/}, $\nabla
g=0$. The geometry whose connection is restricted by the metricity
condition (metric-compatible connection) is called {\it
	Riemann-Cartan geometry\/}.

The connection $\om^{ij}$ determines the parallel transport in the local
Lorentz basis. Because parallel transport is geometric operation it  is
independent of the basis. This property is encoded into PGT via
the so-called {\it vielbein postulate\/}, which implies
\be
\om_{ijk}=\D_{ijk}+K_{ijk}\, ,                               \nn
\ee
where $\D$ is Levi-Civita connection, and
$K_{ijk}=-\frac{1}{2}(T_{ijk}-T_{kij}+T_{jki})$ is the contortion.

\prg{Action and equations of motion.} Lovelock gravity Lagrangian in the first order formulation can be constructed as the linear combination
of the dimensionally continued Euler densities $L_p$, which in $D$ dimensions are defined as
\be
L_p=\ve_{i_1 i_2 \dots i_D} R^{i_1i_2} \dots R^{i_{2p-1} i_{2p}}e^{i_{2p+1}} \dots e^{i_D}.\nn
\ee
In 5$D$ there are three Euler densities and the general form of the action of Lovelock gravity \cite{x1} is
\bea
I=\ve_{ijkln}\int\left(\frac{\a_0}5e^{i}e^{j}e^{k}e^{l}e^{n}+\frac{\a_1}3R^{ij}e^{k}e^{l}e^{n}+\a_2 R^{ij}R^{kl}e^{n}\right)\,. \lab{2.1}
\eea

Variation of the action with respect to vielbein $e^i$
and spin connection $\om^{ij}$ yields the gravitational field equations
\be
\ve_{ijkln}\left(\a_0e^{j}e^{k}e^{l}e^{n}+\a_1R^{jk}e^le^n+\a_2R^{jk}R^{ln}\right)=0\,,\lab{2.2}
\ee
and
\be
\ve_{ijkln}\left(\a_1 e^ke^l+2\a_2R^{kl}\right)T^n=0\,.\lab{2.3}
\ee

\section{Spherically symmetric solution}
\subsection{Ansatz}
\setcounter{equation}{0}
We are looking  for a static solution with $SO(4)$ symmetry, which orbits are three-spheres. The most general metric which fulfills these requirements in Schwarzschild-like coordinates
$x^\m=(t,r,\psi,\th,\vphi)$ is given by
\be
ds^2=N^2dt^2-B^{-2} dr^2-r^2(d\psi^2+\sin^2 \psi d\theta^2+\sin^2 \psi \sin^2 \theta d\vphi^2)\,,\lab{3.1}
\ee
where functions $N$ and $B$ depend solely of $r$, and $r\in [0,\infty)$, $\psi\in[0,\pi)$, $\th\in[0,\pi)$ and $\vphi\in [0,2\pi)$. The metric \eq{3.1} possesses 7 Killing
vectors (see appendix A).

The vielbeins $e^i$ are chosen in a simple diagonal form
\bea
&&e^0=Ndt\,,\qquad e^1=B^{-1}dr\,,\qquad e^2=rd\psi\,,\qquad e^3=r\sin\psi d\th\,,\nn\\
&&e^4=r\sin\psi\sin\th d\vphi\,.
\eea
The most general form of the spin connection compatible with Killing vectors (see appendix A) is given by
\bea\lab{3.3}
&&\omega^{01}=A_0dt+A_1dr\,,\qquad \omega^{02}=A_2d\psi\,,\nn\\
&&\omega^{03}=A_2\sin\psi d\theta\,,\qquad \omega^{04}=A_2\sin\psi \sin\theta d\vphi\,,\nn\\
&&\omega^{12}=A_3d\psi\,,\qquad  \omega^{13}=A_3\sin\psi d\theta\,,\nn\\
&&\omega^{14}=A_3\sin\psi \sin\theta d\vphi\,,\qquad \omega^{23}=\cos\psi d\theta+A_4\sin\psi \sin\theta d\vphi\,,\nn\\
&&\omega^{24}=-A_4\sin\psi d\theta+\cos\psi \sin\theta d\vphi\,,\qquad \omega^{34}=A_4d\psi+\cos\theta d\vphi\,,
\eea
where $A_i$ are arbitrary functions of radial coordinate.
\subsection{Solution}
In the sector with vanishing torsion equations of motion for spherically symmetric ansatz have a well-known solution Boulware-Deser black hole  \cite{x14}, which
exists for the generic choice of action parameters. Another solution, which we  constructed in this paper, possesses  non-vanishing torsion and is given by the following anzatz:
\bea
&&A_0\neq 0\,, \qquad A_1=A_2=A_3=0\,,\qquad A_4\neq 0\nn\\
&&N=B\,.
\eea
By using the adopted anzatz we get that the equations \eq{2.2} reduce to:
\bsubeq
\bea
&&i=0,1:\qquad 2\a_0r^2-\a_1+\a_1 A_4^2=0\,,\lab{x.5a}\\
&&i=2,3,4:\qquad \left(2\a_2-2\a_2A_4^2-\a_1 r^2\right) A_0'+6\a_0r^2+\a_1(A_4^2-1)=0\,.\lab{x.5b}
\eea
\esubeq
The non-vanishing  field equations \eq{2.3}  take the form:
\bsubeq
\bea
&&ij=01: \qquad\a_1r^2+2\a_2A_4^2-2\a_2+4\a_2rA_4A_4'=0\,, \lab{x.6a}\\
&&ij=12,13: \qquad \left(\a_1r^2+2\a_2A_4^2-2\a_2\right)\left(NN'+A_0\right)+2\a_1rN^2=0\,,\lab{x.6b}\\
&&ij=23,24,34:\qquad -2\a_2A_0'+\a_1=0\,.\lab{x.6c}
\eea
\esubeq
From \eq{x.5a} and \eq{x.6c} we get:
\be
 A_4=\sqrt{1-\frac{2 \a_0}{\a_1}r^2}\,,\qquad A_0=\frac{\a_1}{2\a_2}r\,,  \lab{x.7}
\ee
where the integration constant in $A_0$ is taken to be zero for simplicity.
Equation \eq{x.5b} in conjunction with \eq{x.6c} yields to the following  constraint between coupling constants:
\be
\a_1^2-12\a_0\a_2=0\,. \lab{3.5}
\ee
We consequently get that  \eq{x.6a} is identically satisfied, while the \eq{x.6b} takes the form:
$$
NN'+\frac{3N^2}r-\frac{\a_1}{2\a_2}r=0\,,
$$
and can be easily solved for $N$:
\be
N=\sqrt{-\frac{\a_1}{8\a_2}\left(r^2-\frac{r_+^8}{r^6}\right)}\,.\lab{3.4}
\ee
From \eq{3.5} we conclude that the solution exists in the sector different from the Lovelock Chern-Simons gravity. This is exactly
the same fine tuning of parameters found by Canfora et al. in their paper \cite{y2}, where the
 solutions that have the structure of a direct product of a 2D Lorentzian with a 3D Euclidean constant curvature manifold are constructed.

The explicit form of torsion and curvature is given in appendix C. Let us note that both tensorial and antisymmetric part of torsion are non-vanishing
unlike in the case of  the solution found by Canfora et al. \cite{y3}, for which only totally antisymmetric part of torsion is non-vanishing.

Let us now introduce (A)dS radius $\ell$
\be
\frac{\a_1}{8\a_2}=-\frac{\s}{\ell^2}\,,\qquad \s=\pm 1\,.
\ee
By substituting previous relation into \eq{x.7} and \eq{3.4} we get
\be
A_4=\sqrt{1+\frac{4\s r^2}{3\ell^2}}\,,\ N=\sqrt{\s\left(\frac{r^2}{\ell^2}-\frac{r_+^8}{\ell^2r^6}\right)}
\ee
Note that for the solution to describe black hole the following condition must hold
\be
\frac{\a_1}{\a_2}<0\,\Leftrightarrow\, \s=+1\,, \lab{3.8}
\ee
with an \emph{event horizon} located at $r=r_+$.

From the constraint \eq{3.5} it  follows that the sign of the ratio $\frac{\a_0}{\a_1}$ is the same as the sign of $\frac{\a_1}{\a_2}$
\be
{\rm sgn}\left(\frac{\a_0}{\a_1}\right)={\rm sgn}\left(\frac{\a_1}{\a_2}\right)\,.
\ee
 If the ratio is positive expression for $A_4$ implies that we have maximum value of the radial coordinate, so called \emph{cosmological horizon}
\be
r_{0}=\frac{\ell\sqrt 3}2\,.
\ee

While, if the ratio is negative we have no restriction on the value of the radial coordinate, except that it is positive, and in maximally extended space-time goes to infinity. In this case black hole space-time metric is asymptotically AdS.

\prg{Invariants.} From expressions for curvature and torsion, given in appendix C,  we see that quadratic torsional invariant reads
\be
T^i\wedge{}^*T_i=-\frac{12\s}{\ell^2}\left(1-\frac{r_+^8}{r^8}\right)\hat\eps\,,
\ee
which is obviously divergent in $r=0$ for $r_+$ different from zero. Hence, there is a singularity of torsion at $r\ra 0$. Scalar Cartan curvature is constant
\be
R=\frac{16\s}{\ell^2}\,,
\ee
while Riemannian scalar curvature is
\be
\tR=\frac{4\s}{\ell^2}\left(5-\frac{3\s\ell^2}{2r^2}-\frac{3r_+^8}{r^8}\right)\,.
\ee
and is divergent for $r\ra 0$. The quadratic Cartan and Riemannian curvature invariants both vanish
\be
R_{ij}\wedge{}^*R^{ij}=0\,,\quad \tR_{ij}\wedge{}^*\tR^{ij}=0\,.
\ee

We can  conclude that black hole obtained in this article is not of the regular type and that it possesses singularity at $r=0$.
It is worth noting that solution \cite{y3} also possesses singularity of torsion and Riemannian curvature at $r=0$.

Solving equations of motion \eq{2.2} and \eq{2.3} with seven arbitrary functions is an extremely tedious task, which is facilitated by   Mathematica and  xAct packages.

\subsection{Conserved charges}
Conserved charges can be calculated in a number of ways, we decided to make use of Nester's formula \cite{x18}, which application is quite simple in this particular case.
In this section we shall restrict the analysis  to the asymptotically AdS case, which corresponds to the black hole.
The covariant momenta stemming from the Lovelock action \eq{2.1} are given by
\bea
&&\t_i:=\frac{\pd L}{\pd T^i}=0\,,\\
&&\r_{ij}=\frac{\pd L}{\pd R^{ij}}=2\ve_{ijkln}\left(\frac {\a_1}{3}e^ke^l+2\a_2 R^{kl}\right)e^n\,.
\eea
Let us denote the difference between any variable $X$ and
its reference value $\bar X$ by $\D X=X-\bar X$. Reference space-time, in respect to which we measure conserved charges, is given for the zero radius of the event horizon $r_+=0$. Conserved charges $Q_\xi$ associated to the Killing vector $\xi$ are given by quasi-local surface integrals
$$
Q_\xi=\int_{\pd\S}B\,,
$$
where boundary $\pd\S$ is located at infinity. With a suitable asymptotic behavior of the fields, the proper boundary term reads \cite{x18}
\be\lab{3.17}
B=(\xi\inn e^i)\D \t_i + \D e^i(\xi\inn\bt_i)
+\frac 12(\xi\inn\om^i{_j})\D\r_i{^j}
+\frac 12\D\om^i{_j}(\xi\inn\br_i{^j})\, ,
\ee
where $\inn$ denotes contraction.

For solution \eq{3.4}, by making use of the the results of appendix C, we get  the covariant momenta
\bea
&&\r_{01}=\frac{4\left(\a_1^2-12\a_0\a_2\right)}{\a_1}e^2e^3e^4\equiv 0\,,\qquad \r_{02}=-\frac{8\a_1}3 e^1e^3e^4\,,\qquad \r_{03}=\frac{8\a_1}3 e^1e^2e^4\,,\nn\\
&&\r_{04}=-\frac{8\a_1}3 e^1e^2e^3\,,\qquad  \r_{12}=\frac{8\a_1}3e^0e^3e^4-\frac{4\a_1N}{3A_4}e^0e^1e^2\,,\nn\\
&&\r_{13}=-\frac{8\a_1}3 e^0e^2e^4-\frac{4\a_1N}{3A_4}e^0e^1e^3\,,\qquad\r_{14}=\frac{8\a_1}3 e^0e^2e^3-\frac{4\a_1N}{3A_4}e^0e^1e^4\,,\nn\\
&&\r_{23}=0\,,\qquad \r_{24}=0\,,\qquad \r_{34}=0\,.
\eea

From \eq{3.4} we conclude that the connection takes the same form on the background and for $r_+\neq 0$, $\omega^{ij}=\bar{\omega}^{ij}$.
Therefore formula \eq{3.17} takes the following simpler form
$$
B=\frac 12(\xi\inn\om^i{_j})\D\r_i{^j}.
$$

For the seven  Killing vectors $\xi_{(n)}$  (see appendix A)  the conserved charges are given by:
\bea
&&Q_{(0)}=\int_{\pd \S}\om^{01}{}_t\D \r_{01}=0\,,\nn\\
&&Q_{(1)}=\int_{\pd \S}-\cot\psi\sin\th\left(\om^{23}{}_\th\D\r_{23}+\om^{24}{}_\th\D\r_{24}\right)=0\,,\nn\\
&&Q_{(2)}=\int_{\pd\S}\cot\psi\cos\th\cos\vphi\left(\om^{23}{}_\th\D\r_{23}+\om^{24}{}_\th\D\r_{24}\right)\nn\\
&&-\frac{\cot\psi}{\sin\th}\sin\vphi\left(\om^{14}{}_\vphi\D\r_{14}+\om^{23}{}_\vphi\D\r_{23}+\om^{24}{}_\vphi\D\r_{24}+\om^{34}{}_\vphi\D\r_{34}\right)=0\,,\nn\\
&&Q_{(3)}=\int_{\pd\S}\cot\psi\cos\th\sin\vphi\left(\om^{23}{}_\th\D\r_{23}+\om^{24}{}_\th\D\r_{24}\right)\nn\\
&&+\frac{\cot\psi}{\sin\th}\cos\vphi\left(\om^{14}{}_\vphi\D\r_{14}+\om^{23}{}_\vphi\D\r_{23}+\om^{24}{}_\vphi\D\r_{24}+\om^{34}{}_\vphi\D\r_{34}\right)=0\,,\nn\\
&&Q_{(4)}=\int_{\pd\S}\cos\vphi\left(\om^{23}{}_\th\D\r_{23}+\om^{24}{}_\th\D\r_{24}\right)\nn\\
&&-\cot\th\sin\vphi\left(\om^{14}{}_\vphi\D\r_{14}+\om^{23}{}_\vphi\D\r_{23}+\om^{24}{}_\vphi\D\r_{24}+\om^{34}{}_\vphi\D\r_{34}\right)=0\,,\nn\\
&&Q_{(5)}=\int_{\pd\S}\sin\vphi\left(\om^{23}{}_\th\D\r_{23}+\om^{24}{}_\th\D\r_{24}\right)\nn\\
&&+\cot\th\cos\vphi\left(\om^{14}{}_\vphi\D\r_{14}+\om^{23}{}_\vphi\D\r_{23}+\om^{24}{}_\vphi\D\r_{24}+\om^{34}{}_\vphi\D\r_{34}\right)=0\,,\nn\\
&&Q_{(6)}=\int_{\pd\S}\left(\om^{14}{}_\vphi\D\r_{14}+\om^{23}{}_\vphi\D\r_{23}+\om^{24}{}_\vphi\D\r_{24}+\om^{34}{}_\vphi\D\r_{34}\right)=0\,.
\eea

Therefore, we conclude that conserved charges for the black hole with torsion \eq{3.4}  vanish. In particular conserved charge $Q_{(0)}$ which corresponds
to the energy $E$ of the solution vanishes due to {\it specific choice of the parameters} $\a_1^2=12\a_0\a_2$.

\section{Thermodynamics}
\setcounter{equation}{0}
By demanding that Euclidean continuation of the black hole has no conical singularity we obtain the standard formula for the black hole temperature
\be
T=\frac{(N^2)'|_{r=r_+}}{4 \pi}\,.
\ee
In the particular case  of the solution \eq{3.4} we get
\be
T=\frac{2r_+}{\pi\ell^2}\,.\lab{4.2}
\ee
The temperature is positive because solution \eq{3.4} describes black hole iff condition \eq{3.8} is satisfied.
Let us note that this type of relation between temperature and the radius of the event horizon is unusual for  the black holes with spherical horizon. The relation \eq{4.2} is standard in the case of planar black holes (black branes) or black holes in three space-time dimensions.
\subsection{Euclidean action}

By using the equation of motion \eq{2.2}  on-shell Euclidean action takes the form
\be
I_{\rm E}=\varepsilon_{ijklm} \int\left (\frac{2\alpha_1}{3}R^{ij}e^ke^le^m+\frac{4\alpha_0}{5}e^ie^je^ke^le^m\right)\,.
\ee
After substituting the solution \eq{3.4}  we get
\be
I_{\rm E}=\int_{0}^{\beta}dt\int_{r_+}^{\infty}dr\int d\psi d\theta d\vphi\frac{4(\a_1^2-12\a_0\a_2)}{\a_2} r^3 \sin^2 \psi \sin \theta\,,
\ee
where the integration over  time is performed in the interval $[0,\beta:=1/T]$.
By using the constraint on the parameters \eq{3.5}  we conclude that
\be
I_{\rm E}=0.
\ee
From the well-known formula for the entropy
\be
S=(\beta\partial_\beta-1)I_{\rm E}
\ee
we obtain
\be
S=0
\ee
This value of entropy is surprising, but it is not uncommon  for Lovelock black holes, see for instance \cite{x19}, where black holes with zero mass and entropy are obtained.
From Euclidean action we can, also, calculate the energy
\be
E=\pd_\beta I_{\rm E}\,,
\ee
and obtain
\be
E=0
\ee
in accordance with the results of the previous section.
\section{Concluding remarks}

We analyzed static  spherically symmetric solutions of Lovelock gravity in five dimensions. For the generic values of the Lagrangian  parameters the theory possesses a well-known solution, Boulware-Deser black hole, while in the sector $\a_1^2=12\a_0\a_2$ we discovered a new black hole solution with torsion.

We analyzed basic properties of the obtained solution, which torsion possesses non-vanishing tensorial and totally antisymmetric part. The solution has a singularity of torsion
and Riemannian curvature for $r\ra 0$, while the conserved charges, as well as the entropy  vanish.

It worth stressing  that black hole metric is asymptotically AdS, which is a crucial condition for holographic investigation. The solution that describes the space-time which is asymptotically dS,  with the cosmological horizon located at  $r_0=\dis\frac{\a_1}{2\a_0}$, is not a black hole.

An interesting property of the solution in  asymptotically AdS  case is that in semi-classical approximation its entropy is zero. This means that number of micro-states is "small" i.e. it is of order of one instead of the expected $\mathcal{O}(\frac{1}{G_N})$. It would be interesting to see what kind of consequences this result  has on dual interpretation via gauge/gravity duality.
\section*{Acknowledgments}

This work was partially supported by the Serbian Science Foundation  under Grant No. 171031.

\appendix

\section{Killing vectors for metric \eq{3.1}}
\setcounter{equation}{0}

In addition to the $\frac\pd{\pd t}$ Killing vector static and spherically
symmetric  metric \eq{3.1} possesses 6 Killing vectors, due to the $SO(4)$ spherical symmetry. The complete set of
Killing vectors  $\xi^\mu_{(i)}$ of the metric \eq{3.1} is given by:
\bea
&&\xi_{(0)}=\pd_t\,,\nn\\
&&\xi_{(1)}=\cos\th\pd_\psi-\cot\psi\sin\th\pd_\th\,,\nn\\
&&\xi_{(2)}=\sin\th\cos\vphi\pd_\psi+\cot\psi\cos\th\cos\vphi\pd_\th-\frac{\cot\psi}{\sin\th}\sin\vphi\pd_\vphi\,,\nn\\
&&\xi_{(3)}=\sin\th\sin\vphi\pd_\psi+\cot\psi\cos\th\sin\vphi\pd_\th+\frac{\cot\psi}{\sin\th}\cos\vphi\pd_\vphi\,,\nn\\
&&\xi_{(4)}=\cos\vphi\pd_\th-\cot\th\sin\vphi\pd_\vphi\,,\nn\\
&&\xi_{(5)}=\sin\vphi\pd_\th+\cot\th\cos\vphi\pd_\vphi\,,\nn\\
&&\xi_{(6)}=\pd_\vphi\,.
\eea
The independent Killing vectors are $\xi_{(0)}$, $\xi_{(1)}$, $\xi_{(4)}$ and $\xi_{(6)}$, while the others are obtained as their commutators. The invariance conditions of the vielbein under Killing vectors and local Lorentz transformations which  parameters are $\epsilon^i_{\ j}$
\be
\d_0e^i_\mu=L_{\xi}e^i_\mu+\epsilon^i_{\ j}e^j_\mu=0,
\ee
where Lie derivative with respect to $\xi$ is denoted as $L_{\xi}$, gives that the only non-zero parameters of the local Lorentz symmetry are
\be
\epsilon^{23}=-\frac{\sin \theta}{\sin \psi} \ , \ \epsilon^{34}=-\frac{\sin \vphi}{\sin \theta}.
\ee
Using this and the transformation law for spin connection,
\be
\delta_0 \omega^{ij}_\mu=L_{\xi}\omega^{ij}_\mu+\epsilon^i_{\ k}\omega^{kj}_\mu+\epsilon^j_{\ k}\omega^{ik}_\mu=0\,,
\ee
we can derive the most general form of the spherically symmetric spin connection which is given in the main text, formula \eq{3.3}.
\section{Irreducible decomposition of the field strengths}
\setcounter{equation}{0}

We present here formulas for the irreducible decomposition of the PGT
field strengths in a 5D Riemann--Cartan spacetime \cite{x20}.

The torsion 2-form has three irreducible pieces:
\bea
&&{}^{(2)}T^i=\frac{1}{4}b^i\wedge(h_m\inn T^m)\, ,             \nn\\
&&{}^{(3)}T^i=\frac{1}{3}h^i\inn(T^m\wedge b_m)\, ,             \nn\\
&&{}^{(1)}T^i=T^i-{}^{(2)}T^i-{}^{(3)}T^i\, .
\eea
The RC curvature 2-form can be decomposed into six irreducible pieces:
\bsubeq
\be
\ba{ll}
{}^{(2)}R^{ij}=-{}^*(b^{[i}\wedge\Psi^{j]})\, ,
           &{}^{(4)}R^{ij}=\frac 23 b^{[i}\wedge\Phi^{j]}\, ,            \\[3pt]
{}^{(3)}R^{ij}=-\dis\frac{1}{12}X\,{}^*(b^i\wedge b^j)\, ,
           &{}^{(6)}R^{ij}=\dis\frac{1}{20}F\,b^i\wedge b^j\, , \\[9pt]
{}^{(5)}R^{ij}=\dis\frac{1}{3}b^{[i}\wedge h^{j]}\inn(b^m\wedge F_m)\,,
   \qquad  &{}^{(1)}R^{ij}=R^{ij}-\sum_{a=2}^6{}^{(a)}R^{ij}\, .\lab{B.2a}
\ea
\ee
where
\bea
&&F^i:=h_m\inn R^{mi}=\ric^i\, , \qquad F:=h_i\inn F^i=R\, ,    \nn\\
&&X^i:={}^*(R^{ik}\wedge b_k)\, ,\qquad X:=h_i\inn X^i\,.       \lab{B.2b}
\eea
and
\bea
&&\Phi_i:=F_i-\frac{1}{4}b_iF-\frac{1}{2}h_i\inn(b^m\wedge F_m)\,,\nn\\
&&\Psi_i:=X_i-\frac{1}{4}b_i X-\frac{1}{2}h_i\inn(b^m\wedge X_m)\, .
\eea
\esubeq

The above formulas differ from those in  \cite{x20} in two minor
details: the definitions of $F^i$ and $X^i$ are taken with an additional
minus sign, but at the same time, the overall signs of all the irreducible
curvature parts are also changed, leaving their final content unchanged.

\section{Torsion and curvature for the solution \eq{3.4} }
\setcounter{equation}{0}
In this appendix we give values of torsion and curvature for the black hole solution.

\prg{Riemannian connection and curvature.} The non-vanishing components of the Riemannian connection are given by
\bea
&&\tom^{01}=-\frac\s{\ell^2}\left(\frac rN+\frac{3r_+^8}{Nr^7}\right)e^0\,,\qquad \tom^{12}=\frac{N}re^2\,,\qquad\tom^{13}=\frac{N}re^3\,,\nn\\
&&\tom^{23}=\frac{\cot\psi}r e^3\,,\qquad \tom^{14}=\frac{N}re^4\,,\qquad \tom^{24}=\frac{\cot\psi}r e^4\,,\qquad \tom^{34}=\frac{\cot\th}{r\sin\psi}e^4\,.
\eea
Riemannian curvature reads
\bea
&&\tR^{01}=\frac\s{\ell^2}\left(1-\frac{21r_+^8}{r^8}\right)e^0e^1\,,\qquad \tR^{02}=\frac\s{\ell^2}\left(1+\frac{3r_+^8}{r^8}\right)e^0e^2\,,\nn\\
&&\tR^{03}=\frac\s{\ell^2}\left(1+\frac{3r_+^8}{r^8}\right)e^0e^3\,,\qquad \tR^{04}=\frac\s{\ell^2}\left(1+\frac{3r_+^8}{r^8}\right)e^0e^4\,,\nn\\
&&\tR^{12}=\frac\s{\ell^2}\left(1+\frac{3r_+^8}{r^8}\right)e^1e^2\,,\qquad \tR^{13}=\frac\s{\ell^2}\left(1+\frac{3r_+^8}{r^8}\right)e^1e^3\,,\nn\\
&&\tR^{14}=\frac\s{\ell^2}\left(1+\frac{3r_+^8}{r^8}\right)e^1e^4\,,\qquad \tR^{04}=\frac\s{\ell^2}\left(1+\frac{3r_+^8}{r^8}\right)e^0e^4\,,\nn\\
&&\tR^{23}=\frac\s{\ell^2}\left(1-\frac{\s\ell^2}{r^2}-\frac{r_+^8}{r^8}\right)e^2e^3\,,\qquad \tR^{24}=\frac\s{\ell^2}\left(1-\frac{\s\ell^2}{r^2}-\frac{r_+^8}{r^8}\right)e^2e^4\,,\nn\\
&&\tR^{34}=\frac\s{\ell^2}\left(1-\frac{\s\ell^2}{r^2}-\frac{r_+^8}{r^8}\right)e^3e^4\,.
\eea
Riemannian scalar curvature is
\bsubeq
\be
\tR=-\frac{4\s}{\ell^2}\left(-5+\frac{3\s\ell^2}{2r^2}+\frac{3r_+^8}{r^8}\right)\,.
\ee
The quadratic Riemannian curvature invariant vanishes
\be
\tR_{ij}\wedge{}^*\tR^{ij}=0\,.
\ee
\esubeq
\prg{Torsion and its irreducible decomposition.}
The  non-vanishing components of torsion are given by
\bea
&&T^0=\frac{3N}re^0e^1\,,\qquad T^2=\frac{N}re^1e^2+\frac{2A_4}re^3e^4\,,\nn\\
&&T^3=\frac Nre^1e^3-\frac{2A_4}re^2e^4\,,\qquad
T^4=\frac Nre^1e^4+\frac{2A_4}re^2e^3\,.
\eea
The non-vanishing irreducible components of torsion are
\bea
&&{}^{(1)}T^0=\frac{3N}re^0e^1,,\qquad {}^{(1)}T^2=\frac{N}re^1e^2\,,\nn\\
&& {}^{(1)}T^3=\frac{N}re^1e^3\,,\qquad {}^{(1)}T^4=\frac{N}re^1e^4  \,,\nn\\
&&{}^{(3)}T^2=\frac{2A_4}re^3e^4\,,\qquad {}^{(3)}T^3=-\frac{2A_4}re^2e^4\,,\qquad {}^{(3)}T^4=\frac{2A_4}re^2e^3\,.
\eea
The 2nd irreducible component of torsion vanishes as in the case of any solution of Lovelock gravity, excluding Lovelock Chern-Simons \cite{x10}.
Quadratic torsional invariant reads
\be
T^i\wedge{}^*T_i=-\frac{12\s}{\ell^2}\left(1-\frac{r_+^8}{r^8}\right)\hat\eps\,.
\ee
Non-zero components of the (Cartan) curvature are
\bea
&&R^{01}=\frac{4\s}{\ell^2}e^0e^1\,,\qquad
R^{23}=\frac {4\s}{3\ell^2}\frac N{A_4}e^1e^4+\frac {4\s}{3\ell^2}e^2e^3\,,\nn\\
&&R^{24}=-\frac {4\s}{3\ell^2}\frac N{A_4}e^1e^3+\frac {4\s}{3\ell^2}e^2 e^4\,,\qquad
R^{34}=\frac {4\s}{3\ell^2}\frac N{A_4}e^1e^2+\frac {4\s}{3\ell^2}e^3e^4\,.
\eea
Scalar Cartan curvature is constant
\be
R=\frac{16\s}{\ell^2}\,.
\ee
Quadratic Cartan curvature invariant vanishes
\be
R_{ij}\wedge{}^*R^{ij}=0\,.
\ee

\end{document}